\documentclass[12pt]{article}
\usepackage{graphicx}
\newcommand{\beq}{\begin{equation}}
\newcommand{\eeq}{\end{equation}}
\newcommand{\bea}{\begin{eqnarray}}
\newcommand{\eea}{\end{eqnarray}}

\def\la{\mathrel{\mathpalette\fun <}}
\def\ga{\mathrel{\mathpalette\fun >}}
\def\fun#1#2{\lower3.6pt\vbox{\baselineskip0pt\lineskip.9pt
  \ialign{$\mathsurround=0pt#1\hfil##\hfil$\crcr#2\crcr\sim\crcr}}}
\begin{document}
\begin{titlepage}
\begin{flushleft}
       \hfill                      {\tt hep-th/0605nnn}\\
       \hfill                       FIT HE - 06-01 \\
       \hfill                       Kagoshima HE - 06-1 \\
\end{flushleft}
\vspace*{3mm}
\begin{center}
{\bf \Large Holographic Model at Finite Temperature \\

\vspace*{3mm}
with $R$-charge density 
}


\vspace*{5mm}
\vspace*{12mm}
{\large Kazuo Ghoroku\footnote[2]{\tt gouroku@dontaku.fit.ac.jp},
Akihiro Nakamura\footnote[3]{\tt nakamura@sci.kagoshima-u.ac.jp}, 
and Masanobu Yahiro\footnote[4]{\tt yahiro2scp@mbox.nc.kyushu-u.ac.jp}
}\\
\vspace*{8mm}
{\large ${}^{\dagger}$Fukuoka Institute of Technology, Wajiro, 
Higashi-ku}\\
{\large Fukuoka 811-0295, Japan\\}
\vspace*{4mm}
{\large ${}^{\ddagger}$Department of Physics, Kagoshima University, Korimoto 1-21-35,Kagoshima 890-0065, Japan\\}
\vspace*{4mm}
{\large ${}^{\S}$Department of Physics, Kyushu University, Hakozaki,
Higashi-ku}\\
{\large Fukuoka 812-8581, Japan\\}

\vspace*{10mm}

\end{center}
\begin{abstract}
We consider a holographic model of QCD at finite temperature with nonzero
chemical potentials conjugate to $R$-charge densities. A critical surface of
the confinement-deconfinement phase transition is shown for five-dimensional 
charged black hole solution given by Behrnd, Cveti\v{c} and Sabra. On a 
special section of the parameter space, we find an critical curve being similar to the one expected in QCD.
We calculate meson spectra and 
decay constants in the confinement phase of this section
to see their temperature and chemical
potential dependences. 
We could assure generalized GellMann-Oakes-Renner relation and the 
reduction of pion velocity near the critical point.

\end{abstract}
\end{titlepage}

\section{Introduction}
Recently, from the gravity/gauge correspondence \cite{Maldacena:1997re}, 
the probe brane
approach to the gauge theory with flavor quarks
has been largely developed in terms of
the system of $D_p/D_{p+4}$ branes \cite{AdS,GSUY}. 
Inspired by these works, phenomenological, 
5d holographic models have 
been proposed to explain more quantitative
properties of light mesons in a simple setting \cite{EKSS,RP,TB,GMTY,KLS}.
And a simple model 
proposed in \cite{EKSS,RP} is extended to the finite-temperature 
version~\cite{GY}, which
could cover both the confinement and the deconfinement phase.

On the other hand, the thermal gauge theories with chemical potentials 
conjugate to $R$-charges densities have been studied in the framework of the 
gauge/gravity correspondence \cite{Behrndt:1998jd}-\cite{Cai:1998ji}. 
While the charge in such theories is not the one
of QCD, it would be an interesting problem to apply our previous model
\cite{GY}
to these theories and to see the flavor meson properties in a thermal medium
{with chemical potentials conjugate to these charge densities.}

Among the models with $R$-charges, 
we consider here the background configuration which has been found by 
  Behrnd, Cveti\v{c} and Sabra \cite{Behrndt:1998jd} 
as a solution to the equations of motion of the
five-dimensional ${\cal N}=2$ gauged supergravity. This solution is
corresponding to the 
dimensional reduction of the spinning three-brane, and
it is characterized by three chemical potentials and a horizon.
The analyses are given here on a special section of these parameter space
to study the dynamical role of these chemical potentials in the various
quantities related to quark and meson.

In the next section, we set our holographic model of $R$-charge induced 
chemical potential. And a phase diagram, which is very similar to the
one expected in QCD, of this model is given. In the section 3, various
light meson
properties are given and we find many interesting results, especially near
the critical curve of confinement and deconfinement. 
And the summary is given in the final section.

\section{Model setting}
We use the $R$-charged black hole solution to equations of 
motion of the five-dimensional ${\cal N}=2$ gauged supergravity found by
Behrnd, Cveti\v{c} and Sabra \cite{Behrndt:1998jd}. The background for
$R^3$ three space is written as,
\bea
   ds^2_5={ {\cal H}^{-2/3} \over z^2}\left(-f^2(z)dt^2+{\cal H}(dx^{i})^2
+{\cal H}{dz^2\over f^2(z)} \right)
\label{fmet}
\eea
\bea
f^2(z)={\cal H}-({z \over z_{+}})^4 \prod\limits_{i=1}^3 (1+\kappa_i) \,, \quad
{\cal H}=H_1H_2H_3 \,, \quad
H_i=1+\kappa_i({z \over z_{+}})^2 \,, \quad
\kappa_i=q_iz_{+}^2 \,,
\eea
where $q_i\ (i=1\sim 3)$ are the charges of three Abelian gauge groups and 
$z_+$ represents the smallest
value of $f(z)=0$. The radius of AdS${}_5$ is taken as unit. 
The Hawking temperature of the background is given by
\bea
T = 
{2 + \kappa_1 + \kappa_2 + \kappa_3 - \kappa_1 \kappa_2  \kappa_3\over 
2\sqrt{(1+\kappa_1)(1+\kappa_2) (1+\kappa_3)}}\, T_0 
\label{Hawking-T}
\eea
where $T_0=1/(\pi z_+)$ denotes the Hawking temperature in the case of 
the zero chemical potential. Here the chemical potential is defined through 
the three gauge potentials, $A_t^i(z)$, of the gauged supergravity as 
\bea
\mu_i = A_t^i (z)\Biggl|_{z=z_+}= {\sqrt{2 \kappa_i}\over z_{+}(1+\kappa_i)}
 \prod\limits_{l=1}^3
 (1+\kappa_l)^{1/2}\,.  
\eea

\begin{figure}[htbp]
\vspace{.3cm}
\begin{center}
  \includegraphics[width=6.5cm, height=6.5cm]{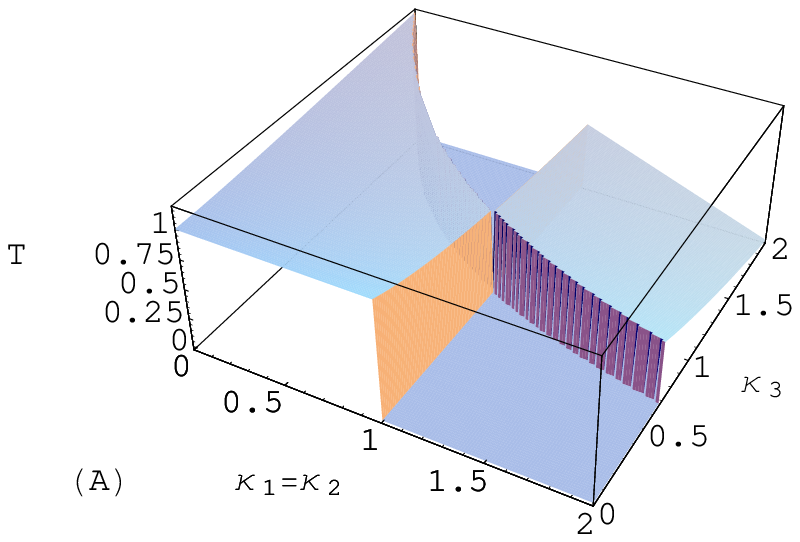}
   \includegraphics[width=6.5cm, height=6.5cm]{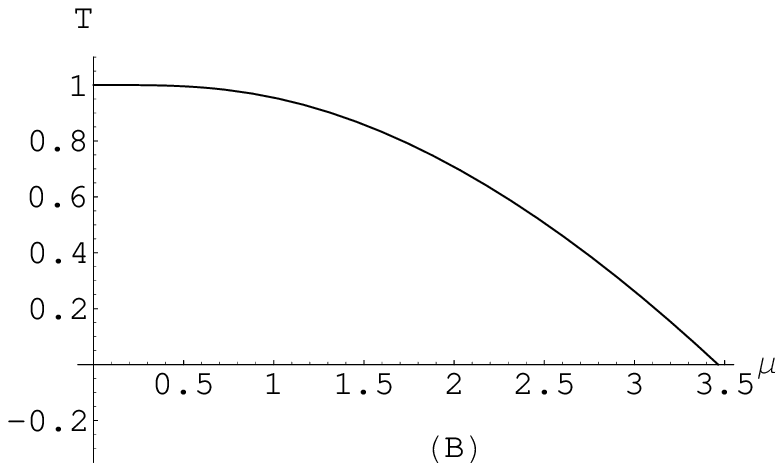}
\caption{(A) Critical surface in the space of 
$((\kappa_1=\kappa_2=)\kappa, \kappa_3, T)$. The surface on the hill represents
the thermodynamic stable region. (B) Critical line obtained 
for $(\mu_1=\mu_2=\mu_3=)\mu$
and $T$. This is the diagonal section of the critical surface in the
Fig.~(A). $T$ and $\mu$ are scaled by $1/\pi z_m$ and $z_m$ respectively.
} \label{QCD-like} 
\end{center}
\end{figure}

\vspace{.5cm}
Since the bulk configuration depends on the gauge potential, 
{we should consider the thermal properties of the system
by considering the Gibbs potential in which the chemical potential 
is fixed. The density of the Gibbs potential $V_G$ is given as
\beq
V_G = -{\pi^2 T_0^4 \over 16G_5} 
 \prod\limits_{i=1}^3 (1+\kappa_i)\,.
\label{Gibbs_density}
\eeq
We find from $V_G$ the condition of thermodynamic stability by demanding
that the system is on the local minimum point of $G$ in $T, \mu_i$
space. This requirement implies the following 
constraint
\footnote
{We notice that the sign of the last term in Eq.~(\ref{ineq})
is opposite to the one of \cite{Son-Sta:2006}. This point is however
important.
}
on $\kappa_i$
\beq
2-\kappa_1-\kappa_2- \kappa_3 + \kappa_1\,  \kappa_2\,   \kappa_3 > 0\,.
\label{ineq}
\eeq
Although various analyses for this system have been given from
the viewpoint of thermodynamics, we consider only the stable region, which
satisfies (\ref{ineq}), in the parameter space. And we
give a critical surface for quark confinement and deconfinement phase 
transition
from a different point of view based on an approximate holographic model.

\vspace{.5cm}
Our idea is as follows.
Usually, the 5d black hole solutions are used to study the high temperature
phase of the dual 4d gauge theories without quark confinement. However,
for the same black hole solutions, 
the low temperature phase with confinement can be realized by introducing
an infrared cutoff
for $z$, $z_m$, such as $z_m<z_+$.
In this case,
the physical region is restricted to the region of $0<z<z_m$. As a result
we can see discrete meson spectra, which are characterized by $z_m$,
and their properties at finite temperature.

On the other hand, for the case of $z_m > z_+$, the physical region is
restricted to $0<z<z_+$ and we find the usual deconfinement background.
Then, in this
formalism, the critical temperature $T_c$ for the 
confinement-deconfinement phase transition is 
given as $T_c=1/(\pi z_m)$ for the case without the chemical potential.
The value of $z_m$ is determined by using some
experimental data of meson mass, for example by the
vector meson ($\rho$ meson) mass, and we obtain $T_c\sim 100\,{\rm MeV}$. 

\vspace{.3cm}
When the chemical potentials are introduced by the model given above, we obtain
in a similar way the critical surface as
\bea
T_c = 
{2 + \kappa_1 + \kappa_2 + \kappa_3 - \kappa_1 \kappa_2  \kappa_3\over 
2\sqrt{(1+\kappa_1)(1+\kappa_2) (1+\kappa_3)}}\, {1 \over \pi z_{m}} \,.
\label{Hawking-Tc}
\eea
This is considered in the 4d
parameter space of temperature and three chemical potentials. In order to
visualize it, we give a typical example
of such a critical surface in a 3d parameter space for $\kappa_1=\kappa_2$
in the Fig.\ref{QCD-like}.
}

The surface on the hill in Fig.~\ref{QCD-like}(A) represents the critical 
surface
on which the black hole solution is thermodynamically stable. Outside
of this hill, the bottom region of the valley, the $R$-charged black hole is
considered to be unstable. So we must consider some other background in this
region. We fortunately find an interesting slice given by the section of 
$\kappa_1=\kappa_2=\kappa_3=\kappa$, from $(T,\kappa)=(1,0)$ to $(0,2)$,
on the critical surface. This section provides
a phase diagram similar to the one expected in QCD.
It is shown in the Fig.~\ref{QCD-like}(B).

Hereafter, we make analyses on the special section, and 
the $\kappa$ ($\mu$) and the $T$ dependence 
of various physical quantities are studied.

\vspace{.3cm}
\section{Meson spectra}
\label{meson}
We consider the 5d meson-action proposed in \cite{EKSS,RP} but use 
the background (\ref{fmet}) to treat the system of finite $T$ and $\kappa$. 
The meson action is 
\beq
S_{\rm meson}=\int d^4x dz\, \sqrt{-g}\, {\rm Tr}\left[-\frac{1}{4g_5^2} 
({L_{MN}L^{MN}}+{R_{MN}R^{MN}}) - {|D_M\Phi|^2} - M^2_\Phi|\Phi|^2\right]\, ,
\label{3-1}
\eeq
where the case of two flavors (${N_f=2}$) is considered; 
notations are the same as in \cite{GY}.
This action is considered as the quark part represented
by the probe brane in the string theory. And the fields on the brane is 
considered to represent meson states composed of a quark-antiquark pair.

\subsection{Scalar field}
The scalar is defined as $\Phi=S\, e^{i\pi^a\tau^a}$ 
and $v(z)\equiv 2 \langle S \rangle$.
The 5d mass of $\Phi$ is $M_{\Phi}^2 =-3$, since the bulk field 
corresponds, in the gauge theory side, 
to an operator with the conformal dimension $\Delta=3$. 
The equation of motion for $v$ is obtained as 
\bea
\left[ \partial_z^2 + 
{z^3 \over f^2}\partial_z({f^2 \over z^3})\cdot  
\partial_z  + {3 {\cal H}^{1/3} \over z^2f^2}
\right]v(z)=0 \,, 
\label{mscalar}
\eea
and $v$ is numerically evaluated in this paper. 
The asymptotic form near $z=0$ 
is $v(z)\sim m_qz+cz^3$, since the background (\ref{fmet}) tends to 
the AdS background there. 
Two constants, $m_q$ and $c$, are identified 
with the quark mass and the chiral condensate, 
respectively.

The $v$ diverges logarithmically at $z=z_+$ 
because of the factor $f$. 
We have confirmed it explicitly for the case of $T > 0$ 
and $\kappa=0$~\cite{GY} with the analytic solution of $v(z)$, 
and numerically 
for the case of finite $\kappa$. 
However, we can evade the dangerous region $z \sim z_+$, 
when $z_+$ is larger than $z_m$ to some extent. 
The present analysis is then focused on the case $z_+ \ga 1.1~z_m$ for 
which the logarithmic divergence is not realized in the physical region 
$z_0 < z < z_m$. 
In analyses shown below, we fix $T$ and see the $\kappa$ dependence of 
physical quantities, since 
the $T$ dependence has already been discussed in the case of 
$\kappa=0$ \cite{GY}.
So the case $z_+ \ga 1.1~z_m$ is realized in 
the lower $\kappa$ region, say $\kappa \la 0.9~\kappa_c(T)$, 
where the critical density 
$\kappa_c(T)$ is a function of $T$ and satisfies 
the condition $z_+=z_m$ for each $T$, that is 
\bea
z_m=
{2 + 3\kappa_c(T) - \kappa_c(T)^3 \over 
2(1+\kappa_c(T))^{3/2} }\, {1 \over \pi T} \,.
\label{kc}
\eea
The treatment of the dangerous region near $\kappa=\kappa_c(T)$ 
will be discussed later.

In the present framework, without any inconsistency, 
we can allow parameters $m_q$ and $c$ to depend on 
$T$ and $\kappa$. However, we do not find a way of determining the $T$ and the 
$\kappa$ dependence, although in real QCD we know that 
the chiral condensate $c$ does depend on $T$. 
This point is the defect of the present model. 
In the present analysis, we simply assume $m_q$ and $c$ are constant 
in the region $\kappa \la 0.9~\kappa_c(T)$. 
This would be a reasonable assumption for $m_q$, but it is just a 
simplification for $c$

The fluctuation of 
$S$, which is defined as $S=v(z)/2+\sigma$, can be
observed as a singlet meson state ($\sigma$). 
Here and hereafter we consider
the static mode, $\partial_i\phi=0$, for any field $\phi$ in order to
derive the mass in a simple way. So the invariant
mass, or pole mass, is defined here as $-\partial_{t}^2\phi=m^2\phi$.  
Then the equation for $\sigma$ is given by adding this 4d mass term 
to Eq.~(\ref{mscalar}) as
\bea
\left[ \partial_z^2 + 
{z^3 \over f^2}\partial_z({f^2 \over z^3})\cdot  
\partial_z  + {3 {\cal H}^{1/3} \over z^2f^2}
+{ {\cal H} \over f^4}m^2
\right]\sigma=0. 
\label{scalareom}
\eea
\\
This equation is independent of $v(z)$. 
The improvement of this defect is postponed to the future.

The discrete mass-spectrum 
is obtained by solving this equation with the boundary conditions,
$\sigma(z) |_{z_0} = \partial_z \sigma(z) |_{z_m} = 0$, where $z_0$
is the UV cutoff which is taken to zero after all. The mass depends on $z_m$, 
$T$ and $\kappa$. First, we determine the value of $z_m$ so as to reproduce 
$\rho$ meson mass at $T=\kappa=0$. The resultant value is 
$1/z_m=0.323$~GeV \cite{GMTY}. It is then possible to estimate 
the $\kappa$ dependence of $\sigma$ meson mass.

\begin{figure}[htbp]
\begin{center}
\voffset=15cm  
  \includegraphics[width=5cm,height=5cm]{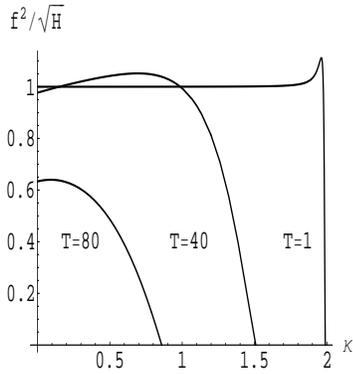}
\caption{ The $\kappa$ dependence of $f^2/\sqrt{\cal H}$ with $T$ fixed. 
Three curves correspond to three cases of $T=1,~40,~80$~MeV, respectively. 
Note that $\kappa_c=1.989$ for the case of $T=1$~MeV, 
1.509 for $T=40$~MeV and 0.858 for $T=80$~MeV. 
}
\label{f-T}
\end{center}
\end{figure}

The $\kappa$ dependence of $m$ is predictable from Eq.(\ref{scalareom}). 
In the equation, the mass term ${m^2{\cal H}/f^4}$ will have a strong 
$\kappa$ dependence through the factor $f^4$, 
if $m$ has no $\kappa$ dependence. However, 
the mass term should have a weaker $\kappa$ dependence, since so do 
the other terms. 
Hence, the strong $\kappa$ dependence of $f^4$ 
is suppressed by that of $m$ in the mass term, indicating that the $\kappa$ 
dependence of $m$ becomes similar to that of ${f^2/\sqrt{{\cal H}}}$.

Fig.~\ref{f-T} shows the $\kappa$ dependence of the factor 
${f^2/\sqrt{{\cal H}}}$ with $T$ fixed. Three curves correspond to three cases 
of $T=1,~40,~80$~MeV, respectively.
The $\kappa$ varies in a region 
$0 \le \kappa < \kappa_c$. 
The phase transition takes place when $\kappa=\kappa_c$. 
For lower $T$ cases such as $T=40$ and 1~MeV, as $\kappa$ increases, 
the factor increases once 
and finally tends to zero in the limit of $\kappa=\kappa_c$.

\begin{figure}[htbp]
\begin{center}
\voffset=15cm
  \includegraphics[width=6cm,height=6cm]{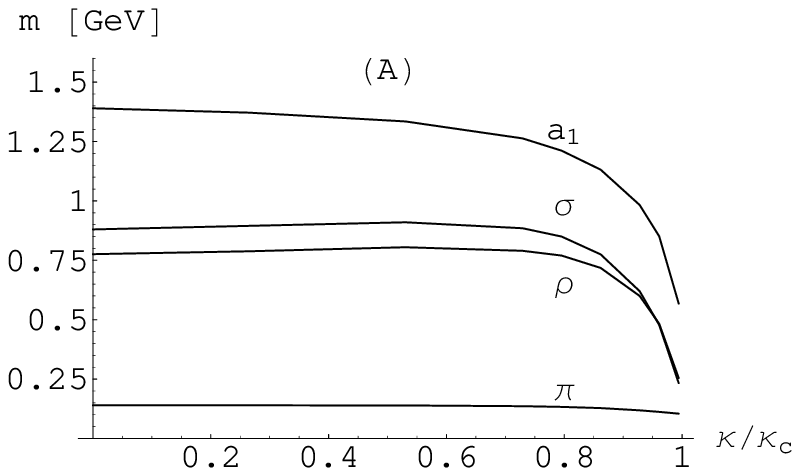}
  \includegraphics[width=6cm,height=6cm]{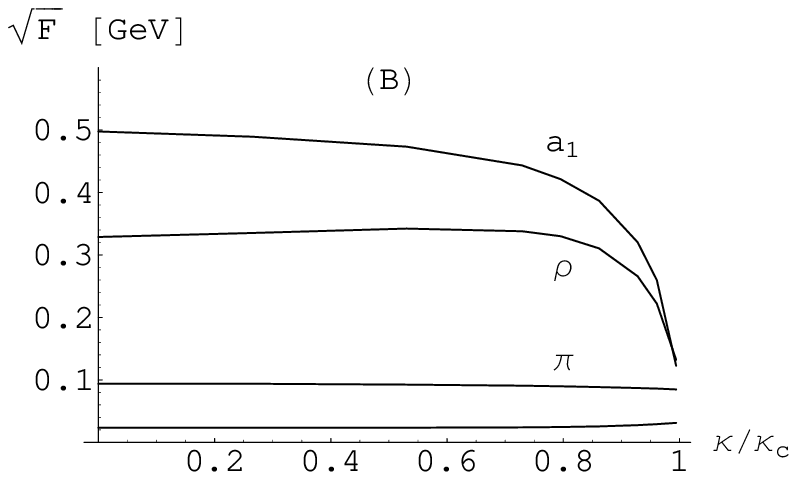}
\caption{ The $\kappa$ dependence 
of masses and decay constants with $T=40$~MeV fixed. 
In (A) the curves show 
$m_{\pi}$, $m_{\rho}$, $m_{\sigma}$ and $m_{a_1}$ from the bottom. 
In (B), the curves denote $F^t_{\pi}/2$, $F^s_{\pi}$, $\sqrt{F_{\rho}}$ 
and $\sqrt{F_{a_1}}$ from the bottom. The $F^t_{\pi}$ increases as 
$\kappa$ goes to $ \kappa_c$, while $F^s_{\pi}$ decreases. 
Other parameters are fixed as $m_q=2.26$~MeV and $c=(0.333 {\rm GeV})^3$.
}
\label{vector40}
\end{center}
\end{figure}

The $\kappa$ dependence of ${f^2/\sqrt{{\cal H}}}$ is compared with 
that of $m$ calculated numerically. The latter is shown in 
Fig.~\ref{vector40} for the case of $T=40$~MeV.
The $\kappa$ dependences of $m$ is found to be similar to 
that of ${f^2/\sqrt{{\cal H}}}$. 
This similarity is true for other $T$. 
Thus, we can see a tendency 
that $m$ decreases as $\kappa$ goes to $\kappa_c$.

For $\sigma$ meson, it is possible to take the Dirichlet 
condition $\sigma(z) |_{z_m}=0$. It 
makes $m$ larger by $\sim 50 \%$, 
but the $\kappa$-dependence of $m$ 
is similar to the case of the Neumann condition.

\vspace{.3cm}
\subsection{Vector meson}
The gauge bosons are separated to
the vector and the axial vector, $V_M$ and $A_M$, and are defined as 
$L_M \equiv V_M + A_M$ and $R_M \equiv V_M - A_M,$ respectively. 

First of all, we consider the vector mesons. The linearized equation 
for the spatial component $V_{i}$ is given as 
\bea
\left[ {m^2{\cal H}\over f^4} + \partial_z^2 
+ {z {\cal H}^{1/3} \over f^2}\partial_z({f^2 \over z {\cal H}^{1/3}})\cdot  
\partial_z \right]V_i=0 \;,
\label{vectoreom}
\eea
where $V_{z}=0$ gauge is taken. 
We adopt the same boundary condition 
$V_i(z)|_{z_0} = \partial_z V_i(z)|_{z_m} = 0$ 
as the case of the scalar meson. 
This equation yields the discrete eigenvalues $m^2=m_n^2$ under 
the boundary condition. 
In order to see the decay constants, it is convenient to expand
$V_i(x,z)$ as $V_i(x,z)=\sum_{n}V_i^{(n)}(x) h_n^V(z)$ for each
mass eigenstate. The spatial component of the
wave-function for each mode is normalized as
\bea
 \int^{z_m}_{z_0}dz {{\cal H}^{2/3} \over z~f^2}(h^V_n(z))^2=1 \,. 
\label{normalization}
\eea
\\
The factor ${\cal H}^{2/3}/(zf^2(z))$ in the integrand is important 
in the sense that it determines the $\kappa$ dependence 
of the decay constant $F_{\rho}$.

Equation (\ref{vectoreom}) is also independent of $v(z)$, 
and the $\kappa$-dependence
of the lowest mass, which is identified with $\rho$ meson mass, 
is shown in Fig.~\ref{vector40}.
The $\rho$-meson mass vanishes or becomes small as $\kappa$ goes to 
$\kappa_c$ because of the mass term $m^2{\cal H}/f^4$.

\subsection{Axial-vector meson and $\pi$-meson}
The linearized equations of motion 
for the axial vector $A_\mu$ and the pion $\pi$ are obtained as 
\bea
\label{axial1}
&&
\left[ {m_a^2{\cal H}\over f^4} + \partial_z^2 
+ {z {\cal H}^{1/3} \over f^2}\partial_z({f^2 \over z {\cal H}^{1/3}})\cdot  
\partial_z - g_5^2{{\cal H}^{1/3}v^2\over z^2f^2} \right]A_{i \perp}=0 .
\\ 
\label{axial0}
&&
\left[ \partial_z^2 + { z \over {\cal H}^{2/3} }\partial_z({ {\cal H}^{2/3} 
\over z })\cdot \partial_z 
- g_5^2{{\cal H}^{1/3}v^2\over z^2f^2} \right] A_{0 \perp} = 0, 
\\
\label{axial2}
&& 
\left[ 
\partial_z^2 + 
{ z \over {\cal H}^{2/3} }\partial_z({ {\cal H}^{2/3} \over z })\cdot \partial_z\right] \varphi 
- g_5^2{{\cal H}^{1/3}v^2 \over z^2f^2} (\pi + \varphi) = 0, 
\\
\label{axial3}
&&
m_{\pi}^2\partial_z \varphi+ 
{g_5^2 { v^2 f^2 \over {\cal H}^{2/3} z^2}}\partial_z \pi=0,  
\eea
where 
$A_\mu$ is decomposed into the transverse and 
the longitudinal part, $A_\mu = A_{\mu \perp} + \partial_\mu \varphi$, 
and $A_{z}=0$ gauge is taken. 
For simplicity, flavor index is neglected and 
$g_5$ is determined from the vector current 
two-point function at $T=\kappa=0$~\cite{EKSS}. 
These equations are solved numerically under the boundary conditions, 
$A_{\mu \perp}(z_0)=\partial_zA_{\mu \perp}(z_m)=0$ and  
$\varphi(z_0)=\partial_z\varphi(z_m)=\pi(z_0)=0$.

Differently from the case of vector and $\sigma$ mesons,
the quantities of axial-vector and pion 
depend on five parameters $m_q$, $c$, $z_m$, $T$ and $\kappa$ 
through $v(z)$, $f(z)$ and ${\cal H}$ 
as seen in Eqs. (\ref{axial1})-(\ref{axial3}).
For the consistency between the vector and axial-vector meson sectors, here 
we take the same $z_m$ as in the vector meson sector. 
Parameters, $m_q$ and $c$, are determined to 
reproduce the experimental values, $\bar{m}_\pi$ and $\bar{F}_\pi$, 
of $m_\pi$ and $F_\pi$ at $T=\kappa=0$; 
the resultant values are $m_q=2.26$~MeV and $c=(0.333~{\rm GeV})^3$. 
This parameter set reproduces masses and decay constants of 
$\pi$, $\sigma$, $\rho$, $a_1$ mesons at $T=\kappa=0$ within $\sim 10 \%$ 
error~\cite{EKSS,GSUY}. 
The present model is also successful in reproducing the $T$ 
dependences of masses and decay constants of $\pi$ and $\rho$ mesons 
calculated with the model-independent methods such as 
lattice QCD calculation~\cite{Karsch2} and 
the chiral perturbation theory~\cite{Toublan}.

In principle, the chiral condensate $c$ depends on $T$ and $\kappa$, 
but we simply assume that $c$ is independent of $T$ and $\kappa$. 
As shown in Fig.~\ref{vector40}, 
masses of axial-vector meson ($a_1$) and pion also become small 
as $\kappa$ goes to $\kappa_c$.

\subsection{Decay constants}
The decay constants are obtained from the wave functions as~\cite{EKSS,GY} 
\bea
F_{a_n}^2 = \frac{1}{g_5^2} \left[ 
\left. \frac{d^2h^{A}_n}{dz^2} \right|_{z_0} \right]^2, \quad
({F^{t,s}_\pi})^2 = - \frac{1}{g_5^2} 
  \left. \frac{\partial_z {A}^{(0)}_{0,i \perp}}{z} \right|_{z_0} ,  
\nonumber
\label{F-pi}
\eea
where 
$A_{i \perp}(x,z) = \sum_{n \ge 1} \alpha^{(n)}_{i}(x) h_n^A(z)$ 
and 
$h_n^A(z)$ is normalized as $h_n^V(z)$ given in (\ref{normalization}).
Furthermore, $A^{(0)}_{0 \perp}$ ($A^{(0)}_{i \perp}$) is 
the solution to 
Eq. (\ref{axial0}) (Eq. (\ref{axial1}) with $m_a^2=0$), 
satisfying 
$A^{(0)}_{0,i \perp}(z_0)=1$ and 
$\partial_z A^{(0)}_{0,i \perp}(z_m)=0$, 
and $F^{t,s}_\pi$ are the timelike and spatial components 
of the pion decay constant. Obviously, 
$F^{t}_\pi$ and $F^{s}_\pi$ are different from each other at finite $T$ or 
$\kappa$, 
while $F^t_\pi=F^s_\pi$ at $T=\kappa=0$.

We can see from Figs.~\ref{vector40} that 
the pion decay constants decreases as $\kappa$ increases. Eventually, we find 
${m(\kappa)/m(0)}\sim {F(\kappa)/F(0)}$. 
Note that $F^t_{\pi}$ goes up slowly as $\kappa$ increases. 
This is an exception and important 
in determining the $\kappa$ dependence of pion velocity $v_{\pi}$, 
as mentioned later.

\subsection{GOR relation and pion velocity}

\begin{figure}[htbp]
\begin{center}
\voffset=15cm
  \includegraphics[width=6cm,height=6cm]{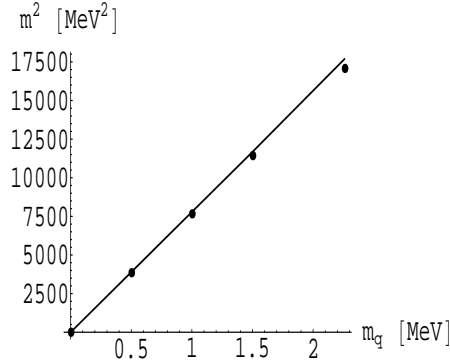}  
\caption{The $m_q$ dependence of $m_{\pi}^2$. 
Other parameters are fixed at $T=40$~MeV, 
$\kappa=0.8~\kappa_c$, $\kappa_c=1.509$ and $c=(0.333~{\rm GeV})^3$. 
The solid line corresponds to $m_{\pi}^2(\kappa)$ and 
the closed circles to $2m_q c/{F^t_\pi}^2(\kappa)$. 
}
\label{GOR}
\end{center}
\end{figure}

In real QCD, a generalized GOR relation, $m_\pi^2{F_\pi^t}^2 = 2m_q c$, 
is expected 
for finite $T$~\cite{Pisarski,Toublan}. 
In the previous work~\cite{GY}, we confirmed that the relation is realized 
in our model for the case of $T > 0$ and $\kappa=0$. 
It is then of interest whether the relation persists even for finite $\kappa$. 
This is tested by calculating $m_{\pi}$ and ${F^t_\pi}$ numerically. 
Fig.~\ref{GOR} shows the $m_q$ dependence of 
$m_{\pi}^2$ and $2m_q c/{F^t_\pi}^2$ by 
the solid line and the closed circles, respectively. 
Here the case of $T=40$~MeV and $\kappa=0.8~\kappa_c$ 
is taken as an example. 
The $m_{\pi}^2(T)$ shown by the solid line 
tends to zero as $m_q$ decreases, 
as a reflection of the Nambu-Goldstone theorem at finite $T$ and $\kappa$.
Comparing the solid line and the closed circles, one can see that 
the generalized GOR relation is satisfied.

The pion velocity $v_{\pi}$ in the thermal midium with finite $\mu$ 
can be estimated analytically in the chiral limit, $m_q=0$, from 
$F^t_\pi$ and $F^s_\pi$ as $v_{\pi} = F^s_\pi/F^t_\pi$~\cite{Pisarski,SS}. 
In the outer region of the critical line, i.e. when $\kappa > \kappa_c$, 
we expect the restoration of chiral symmetry.
So we can set as $v(z)=0$ for any $\kappa$ near $\kappa_c$, 
at least when the transition is the second order. 
This $v$ has no divergence, even when $z_+=z_m$. We can then 
consider the limit $\kappa \to \kappa_c(T)$. 
The solutions of Eqs.(\ref{axial1}) and (\ref{axial0})  are given, 
under the boundary conditions $A_{i,0}|_{z=0}=1$, as
\bea
 A_i&=&1+b_i \int_0^z dz_1{z_1{\cal H}^{1/3}\over f^2} \, ,
\\
A_0&=&1+b_0 \int_0^z dz_1{z_1 \over {\cal H}^{2/3}} \, .
\eea
The constants $b_{0,i}$ are determined by the boundary
conditions, $\partial_zA_{i,0}|_{z=z_m}=\epsilon$, 
and the limit $\epsilon\to 0$ is taken after obtaining $v_{\pi}$. 
We then get 
\begin{equation}
  {v_{\pi}}^2={f^2(z_m) \over {\cal H}(z_m)} = 
1-({z_m\over z_+})^4 
\;. 
\end{equation}
We are now considering the limit of
$z_+\to z_m$, so $v_{\pi}^2$ approaches to zero linearly with respect to
the difference $z_+ - z_m$. 
This implies ${v_{\pi}}^2\propto 1-{T\over T_c}$ for the
path of $z_+\to z_m$ with $\kappa$ fixed and 
that ${v_{\pi}}^2\propto \kappa_c-\kappa$ for the path with $T$ fixed. 
Then we can assure the critical exponent $\nu=1$ for both cases.

Finally, in the chiral limit, we discuss the behavior of 
the chiral condensate $c$ near $\kappa=\kappa_c(T)$. 
When $c$ is finite, $v(z)$ diverges at $z=z_+$, 
although the divergence is logarithmic and then very weak. 
When $\kappa=\kappa_c(T)$, the divergence is realized 
in the present model which considers the region $z_0 < z < z_m$. 
As mentioned above, the $\kappa$ dependence of $c$ is not determined within 
the present model. However, if the deconfinement transition and the 
chiral one take place simultaneously at $\kappa=\kappa_c(T)$ and 
the chiral transition is the second order, we might evade the logarithmic 
divergence of $v(z)$, since $c$ is considered to tend 
to zero more strongly as $\kappa$ approaches $\kappa_c(T)$. 
In this situation, we can evaluate $v_{\pi}$ at $\kappa=\kappa_c(T)$.

\vspace{.3cm}
\section{Summary}

In this paper, we constructed a holographic model of gauge theory at finite 
temperature with non-zero chemical potential ($\mu$) 
conjugate to $R$-charge denisty ($\kappa$), 
introducing the five-dimensional 
$R$-charged black hole solution of Behrnd, Cveti\v{c} and Sabra. 
The gauge theory thus constructed has two phases, 
confinement and deconfinement ones.  
The phase diagram of the gauge theory is richer than that of QCD in 
the sense that the gauge theory has three chemical potentials conjugate to 
$R$-charge densities while QCD has only one chemical potential conjugate to 
the baryon number density. In the three-dimensional phase diagram of 
the gauge theory, a special section is similar to the one expected in QCD.

The present model makes it possible to evaluate 
meson masses and decay constants nonperturbatively even in the thermal system 
with non-zero chemical potential. 
We then calculated these quantities 
in the confinement phase of the special section and 
found that these quantities decrease as $R$-charge density $\kappa$ 
increases with $T$ fixed. 
Furthermore, we confirmed that the Nambu-Goldstone theorem and 
the generalized GellMann-Oakes-Renner relation 
persist in the thermal system with finite chemical potential. 

We estimated the pion velocity as $v_{\pi}=0$ on 
the critical line of the phase diagram, 
assuming that the deconfinement phase transition and the second-order 
chiral one take place simultaneously. 
The present result is consistent with the result of 
Son and Stephanov~\cite{SS} based on the sigma model, 
but somewhat deviates from 
the result of Ref. \cite{SH}. However, the present result supports 
the statement that 
the measured pion velocity 0.65~\cite{Cramer}, deduced from the pion spectra 
observed by STAR~\cite{STAR} at RHIC, would be a signal of QCD phase 
transition.

The present model has two defects. 
One is that the profile function 
$v(z)$ diverges on the critical line, 
although the divergence is logarithmic and then very weak. 
The other is that the $T$ and the $\kappa$ dependence of $c$ are 
not determined within 
the present model. As for the chiral limit, however, 
if the deconfinement phase transition 
and the second-order chiral one take place simultaneously, 
we can expect that $c$ tends to zero as $T$ and $\kappa$ approach the 
critical line.  For such a vanishing $c$, we can expect that 
$v(z)$ does not diverge on the critical line and then that 
the present model becomes reliable even near the line. 
So it is quite interesting to find a way of determining the 
$T$ and the $\kappa$ dependence of $c$ 
in the present holographic model or its extension.

Finally we comment on the bold procedure, the hard IR cutoff, which
is essential in the present model. This cutoff seems to be bold, but we
could understand it very naturally. The action $S_{\rm meson}$, Eq.(\ref{3-1}),
is regarded as
the effective 5d action for the probe brane(s). In the phase of chiral
symmetry breaking, we find a minimum value of $r$, $r_m$, for the probe brane
by solving its profile function \cite{AdS}, so the
theory on the brane should be restricted to the region $r_{m}<r$. This $r_m$
could be identified with the IR cutoff introduced in the present model.

\vspace{.3cm}
\section*
{\bf Acknowledgments}
This work has been supported in part by the Grants-in-Aid
for Scientific Research (13135223)
of the Ministry of Education, Science, Sports, and Culture of Japan.




\begin{thebibliography}{99}
\bibitem{Maldacena:1997re}
  J.~M.~Maldacena,
   ``The large N limit of superconformal field theories and supergravity,''
  %
  Adv.\ Theor.\ Math.\ Phys.\  {\bf 2}, 231 (1998)
  [Int.\ J.\ Theor.\ Phys.\  {\bf 38}, 1113 (1999)]
  [arXiv:hep-th/9711200].

  S.~S.~Gubser, I.~R.~Klebanov and A.~M.~Polyakov,
   ``Gauge theory correlators from non-critical string theory,''
  %
  Phys.\ Lett.\ B {\bf 428}, 105 (1998)
  [arXiv:hep-th/9802109].

  E.~Witten,
   ``Anti-de Sitter space and holography,''
  %
  Adv.\ Theor.\ Math.\ Phys.\  {\bf 2}, 253 (1998)
  [arXiv:hep-th/9802150].
\bibitem{AdS}
  A.~Karch and E.~Katz, 
  JHEP {\bf 0206}, 043(2003).
  M.~Kruczenski, D.~Mateos, R.C.~Myers and D.J.~Winters, 
  JHEP {\bf 0307}, 049(2003).
  JHEP {\bf 0405}, 041 (2004).
  J.~Babington, J.~Erdmenger, N.~J.~Evans, Z.~Guralnik and I.~Kirsch,
  Phys.\ Rev.\ D {\bf 69}, 066007 (2004). 
  N.~J.~Evans and J.~P.~Shock,
  Phys.\ Rev.\ D {\bf 70}, 046002 (2004). 
  C.~Nunez, A.~Paredes and A.V.~Ramallo, 
  JHEP {\bf 0312}, 024(2003). 
 T. Sakai and S. Sugimoto, Prog.Theor.Phys.113(2005)843-882;
  K. Ghoroku and M. Yahiro, 
  Phys.\ Lett.\ B {\bf 604}, 235 (2004). 

\bibitem{GSUY} K. Ghoroku, T. Sakaguchi, N. Uekusa and M. Yahiro,
 Phys. Rev. D71(2005)106002. 

\bibitem{EKSS}
  J.~Erlich, E.~Katz, D.~T.~Son and M.~A.~Stephanov,
   hep-ph/0501128.
\bibitem{RP}
  L.~Da Rold and A.~Pomarol,
   Nucl.\ Phys.\ B {\bf 721}, 79 (2005); 
   hep-ph/0510268.
\bibitem{TB}
  G.~F.~de Teramond and S.~J.~Brodsky,
  Phys.\ Rev.\ Lett.\  {\bf 94}, 201601 (2005). 

\bibitem{GMTY} K. Ghoroku, N. Maru, M. Tachibana and M. Yahiro,
 hep-ph/0510334.

\bibitem{KLS} E.~Katz, A. Lewandowski and M.D. Schwartz,
  hep-ph/0510388.


\bibitem{GY} 
  K. Ghoroku and M. Yahiro,  [arXiv:hep-th/0512289].


\bibitem{Behrndt:1998jd}
  K.~Behrndt, M.~Cvetic and W.~A.~Sabra,
  ``Non-extreme black holes of five dimensional N = 2 AdS supergravity,''
  Nucl.\ Phys.\ B {\bf 553}, 317 (1999)
  [arXiv:hep-th/9810227]. 

\bibitem{GG} 
  R. V. Gavai and S. Gupta Phys.\ Rev.\ D{\bf 68}, 034506(2003) 
[arXiv:hep-lat/0303013].

\bibitem{Son-Sta:2006}
D.~T.~Son and A.~O.~Starinets, [arXiv:hep-th/0601157].
\bibitem{Kovtun:2003wp}
  P.~Kovtun, D.~T.~Son and A.~O.~Starinets,
   ``Holography and hydrodynamics: Diffusion on stretched horizons,''
  %
  JHEP {\bf 0310}, 064 (2003)
  [arXiv:hep-th/0309213].


\bibitem{Gubser:1998jb}
  S.~S.~Gubser,
  ``Thermodynamics of spinning D3-branes,''
  Nucl.\ Phys.\ B {\bf 551}, 667 (1999)
  [arXiv:hep-th/9810225].




\bibitem{Chamblin:1999tk}
  A.~Chamblin, R.~Emparan, C.~V.~Johnson and R.~C.~Myers,
  ``Charged AdS black holes and catastrophic holography,''
  Phys.\ Rev.\ D {\bf 60}, 064018 (1999)
  [arXiv:hep-th/9902170].




\bibitem{Cvetic:1999ne}
  M.~Cvetic and S.~S.~Gubser,
  ``Phases of $R$-charged black holes, spinning branes and strongly coupled
  gauge theories,''
  JHEP {\bf 9904}, 024 (1999)
  [arXiv:hep-th/9902195].


\bibitem{Cai:1998ji}
  R.~G.~Cai and K.~S.~Soh,
   ``Critical behavior in the rotating D-branes,''
  %
  Mod.\ Phys.\ Lett.\ A {\bf 14}, 1895 (1999)
  [arXiv:hep-th/9812121].


\bibitem{STAR}
J. Adams et al. [STAR Collaboration], Phys. Rev. Lett. {\bf 92}, 112301(2004).

\bibitem{Toublan}
D. Toublan, Phys. Rev. {\bf D56}, 5629(1997). 

\bibitem{Karsch2}
  F. Karsch, Nucl. Phys. Proc. Suppl. {\bf D83}, 14(2000). 

\bibitem{Pisarski}
R.D. Pisarski and M. Tytgat, Phys. Rev. {\bf D54}, 2989(1996). 

\bibitem{SS}
D. T. Son and M. A. Stephanov, Phys. Rev. Lett. {\bf 88}, 202302 (2002).

\bibitem{SH} 
M. Harada, M. Rho and C. Sasaki, hep-ph/0506092.

\bibitem{Cramer}
J. G. Cramer, G. A. Miller, J. M. S. Wu and J. H. S. Yoon, Phys. Rev. Lett. 
{\bf 94}, 102302(2005).

\end{thebibliography}
\end{document}